\documentclass{iau} 
\usepackage{graphicx}
\usepackage{amsmath}
\usepackage{url}
\usepackage{hyperref}
\usepackage{gensymb}
\usepackage{caption}
\usepackage{comment}
\usepackage{floatrow}
\usepackage[rightcaption]{sidecap}
\sidecaptionvpos{figure}{c}

\title[Patchy spectra of pulsar individual pulses] 
{On the patchiness of the individual pulse spectra at the very low radio frequencies}

\author[Song, Kondratiev \& Bilous]   
{Xiaoxi Song$^1$, Vladislav Kondratiev$^{2}$, \and Anna Bilous$^3$}

\affiliation{$^1$Department of Physics and Astronomy, University College London, \\ Gower Street, London, UK, WC1E 6BT, email: {\tt xiao.song.13@ucl.ac.uk} \\[\affilskip]
$^2$ASTRON, the Netherlands Institute for Radio Astronomy, \\ Postbus
2, 7990 AA Dwingeloo, The Netherlands \\[\affilskip]
$^3$Anton Pannekoek Institute for Astronomy, University of Amsterdam, \\ Science Park 904, 1098 XH Amsterdam, The Netherlands \\[\affilskip]
}

\pubyear{2017}
\volume{337}  
\jname{Pulsar Astrophysics -Â­ The Next 50 Years}
\editors{P. Weltevrede, B.B.P. Perera, L. Levin Preston \& S. Sanidas, eds.}

\begin{document}

\maketitle
\vskip -3mm
\begin{abstract}
We have used sensitive LOw Frequency ARray (LOFAR) observations of PSR B0809+74 at 15--62 MHz to study the anomalously intensive pulses, first reported by \cite[Ulyanov et al. (2006)]{Ulyanov2006} at 18--30\,MHz. Similarly to Ulyanov et al., we found that the spectra of strong pulses consist of distinct bright patches. Moreover, these spectral patches were spotted to drift upwards in frequency over the course of several pulse sequences. We established that this drift is not pulsar-intrinsic, but is caused by the broadband  $\sim$20\,second-long enhancements of recorded signal, which influenced the dispersed tracks of several pulses at once. We speculate on the cause of such enhancements (i.e. propagation or telescope-related) and the ramifications they bring to the single-pulse studies at the very low radio frequencies. Depending on the origin, the phenomenon may also affect the analysis of highly dispersed single pulses at higher radio frequencies, e.g. Fast Radio Bursts.

\end{abstract}

The properties of individual pulses of pulsar radio emission are directly related to the emission microphysics and thus can help constraining the elusive pulsar radio emission mechanism. So far, single pulses have been studied mostly at frequencies around 1--2\,GHz, but with the new generation of low-frequency radio telescopes the spectral coverage can be extended to the decameter wavelengths. Recently, \cite[Ulyanov et al. (2006)]{Ulyanov2006} reported on the discovery of anomalously intensive pulses (AIPs) at 18--30\,MHz, which had energies similar to giant pulses (GPs). Contrary to GPs, known only for a few young or millisecond pulsars, AIPs were coming from the ``ordinary" non-recycled sources.

\begin{figure}[htp]
\begin{center}
 \includegraphics[width=\textwidth]{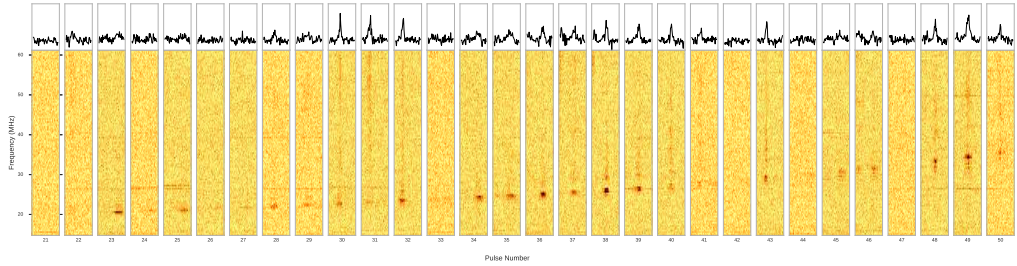} 
 \caption{Spectra of 30 pulses with the coherent frequency drift. Only the on-pulse region is shown (20\% of the pulse phase), with band-integrated profiles on the top row.}
   \label{fig:supersequence}
\end{center}
\end{figure}

We have observed several AIP pulsars with LOFAR's (Low Frequency ARray's) Low Band Antennas using ``Superterp" stations at 15--62\,MHz and found the MHz-wide spectral structure in bright single pulses from a number of observations. The full analysis will be reported in Song et al. (in prep.), and here we focus on one pulse sequence from the observations of PSR~B0809+74, where the bright frequency patches appeared to be coherently drifting upwards in frequency over the course of $\sim 50$ pulses (Fig.~\ref{fig:supersequence}, 30 pulses are shown). To our knowledge, such systematic frequency drift has not been reported in the literature before. 

The careful examination of dynamic spectra revealed that the drifting patches followed the inverse-dispersion track, thus
firmly pointing to the pulsar-extrinsic cause of the observed phenomenon. The appearance of patches was correlated with the apparent $\sim20$\,second-long broadband increase of the background intensity levels, and flattening the 2-D (time and frequency) baseline revealed that it was coincident with broadband increase of the signal-to-noise (S/N) ratio. With DM of $\approx 5.7$\,pc\,cm$^{-3}$, the dispersive delay across the observing band is 
$\approx 107$\,s (much longer than 1.3-s pulsar period), thus the region of 

\thisfloatsetup{
  capposition=beside,
  capbesideposition={top,right},
  capbesidewidth=.45\textwidth,
  justification=raggedright
}
\begin{figure}[H]
\floatbox{figure}[0.5\textwidth]{\caption{Schematic demonstration of the frequency drift origin. The pink vertical stripe represents the region of broadband S/N increase and the black lines mark dispersed tracks of the individual pulses. The term ``sensitivity" on the y-axis of the bottom-left plot is used broadly to indicate either signal enhancement (e.g. due to scintillation) or gain variation (e.g. instrumental).}\label{fig:demo}}
{\includegraphics[trim=15mm 10mm 0 8mm,width=0.55\textwidth]{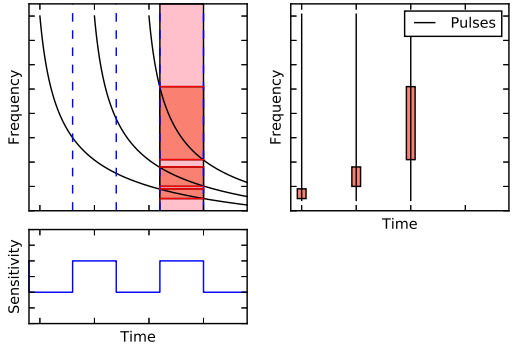}}
\end{figure}
\noindent S/N increase will affect several pulses passing through it, and after de-dispersion, will manifest itself as patches in the spectra of individual pulses (Fig.~\ref{fig:demo}). The spectral width affected will be larger at higher frequencies, just as observed.

Several reasons may serve as an explanation for the S/N increase: 

{\underline{\it Scintillation on ISM or IPM}. Unlikely. The decorrelation bandwidth for diffractive scintillation is too small (few kHz) and the time scale of refractive scintillation is too long (min--hr). For scintillation on IPM or solar wind the time scale is too short (1--2\,s), and PSR B0809+74 is at high ecliptic latitude, so it is much less affected by solar wind.

{\underline{\it Ionospheric scintillation}}. Very likely. As, in general, the time scale is comparable ($\sim 10$--100\,s, \cite[Loi et al. 2016]{Loi2016}). In the case of our observation of PSR~B0809+74 we also see accompanying variation of the baseline, which can not be explained by ionospheric scintillation.

{\underline{\it Beam wandering due to refraction in the ionosphere or IPM}}. Possible. Beam wandering can be larger than 10 arcmin depending on heliospheric distance, a substantial fraction of the LOFAR Superterp beam \cite[(van Haarlem et al. 2013)]{vanHaarlem2013}.

{\underline{\it Instrumental issues}}. Possible. For instance, the subset of subbands from one-two stations can drop out, causing sensitivity changes.

The observed broadband S/N variations have potentially serious implications for the low-frequency single-pulse analysis, biasing the spectra, energy distributions single-pulse DM measurements, and so on. Depending on the cause, similar effect may appear at higher frequencies, biasing the spectra of highly dispersed pulses, e.g. Fast Radio Bursts.

\end{document}